\documentclass{ws-procs11x85}

\usepackage{multicol}
\def\Journal#1#2#3#4{{#1} {\bf #2}, #3 (#4)}

\makeindex
\begin{document}

\title{Confinement and fat-center-vortices model}

\author{Sedigheh Deldar}

\address{Department of Physics, University of Tehran
\\E-mail:sdeldar@khayam.ut.ac.ir}


\maketitle\abstract{
In this paper I review shortly potentials obtained for SU(2), SU(3) and SU(4)
static sources from fat-center-vortices model. Results confirm the confinement
of quarks in all three gauge groups. Proportionality of string tensions with
flux tube counting is better than Casimir scaling especially for SU(4).} 

\begin{multicols}{2}

\baselineskip=13.07pt
\section{Introduction}

Today there is no doubt about the confinement of quarks which claims that
quarks are not free in the nature and all we see are in color singlet states.
In other words the potential between quarks is linear and increases with
distance 
\begin{equation}
V(r)=-\frac{A}{r} + Kr + c
\end{equation} 
where the first term is Coulombic and $K$ is the coefficient of the linear part 
of the potential and is called string tension.  
Many numerical calculations in SU(2), SU(3) and SU(4) for fundamental
and higher representations \cite{fund,Bali} confirm the confinement of quarks in QCD.
Also there are some phenomenological models which explain the confinement in terms 
of gauge field configurations included instantons, merons, abelian monoploes 
and center vortices. QCD vacuum is assumed to be constructed from these gauge
field configurations. In this paper I review very shortly fat center vortices 
model and take a look at the results obtained using this model for SU(2) 
\cite{Gren}, SU(3) \cite{Deld} and SU(4) \cite{Rafi}. I discuss about "Casimir
scaling" and flux tube counting, the two possible candidates for explaining the
behavior of string tension of quarks in higher representations at intermediate 
distances.

\section{Fat-center Vortices model}

The center vortex theory was introduced by t'Hooft \cite{Hoof} in the late 
1970's. The theory is able to explain quark confinement by describing the
interaction between a center vortex  - which is indeed a topological field
configuration - and a Wilson loop. Potential between two quarks can be measured 
by studying the Wilson loop at large $T$:
\begin{equation}
W(R,t) \simeq \exp^{-V(R)T}
\end{equation}
$W(R,T)$ is the Wilson loop as a function of $R$, the
spatial separation of quarks, and the propagation time $T$, and $V(R)$
is the gauge field energy associated with the static quark-antiquark
source. Briefly, the vortex theory states that the area law falloff 
for large $T$ of Wilson loop is due to fluctuations in the number of center
vortices linking the loop. 
Potential between adjoint quarks is zero at large distances. This is because 
adjoint quarks are screened by gluons at this distance. Therefore center 
vortices do not have any effect on the Wilson loop of higher representations.
But before the onset of screening 
there exists a region where the potential between two sources is linear.
Fat center vortices introduced by Greensite {\it et al.} \cite{Gren} has been 
able to predict this linearity of the potential for all representations. Based 
on this model the vortex is thick enough 
such that its core somewhere overlaps the perimeter of the Wilson loop.
More details about this improved vortex theory can be found in the paper of
reference \cite{Gren}. Here I only introduce the Wilson loop obtained from
this model and discuss results for SU(2), SU(3) and SU(4) quarks.

The average Wilson loop is given by:
\begin{equation}
<W(C)> = \prod_{x} \{ 1 - \sum^{N-1}_{n=1} f_{n} (1 - Re {\cal G}_{r}
          [\vec{\alpha}^n_{C}(x)])\},
\label{WC}
\end{equation}
$x$ is the location of the center of the vortex and A is the area of the
loop C and ${\cal G}_{r}$ is:
\begin{equation}
{\cal G}_{r}[\vec{\alpha}] = \frac{1}{d_{r}} Tr \exp[i\vec{\alpha} . \vec{H}].
\label{gr}
\end{equation}
where $d_{r}$ is the dimension of the representation and
$\{H_{i},i=1,2,...,N-1\}$  are generators of the group. $f$ is the
probability that any given unit is pierced by a vortex and depends on what
fraction of the vortex core is enclosed by the Wilson loop. It also depends
on the shape of the loop and the position of the center of the vortex in the
plane of loop C relative to the perimeter. $\alpha_{R}(x)$, the flux 
distribution of the vortex can be chosen in such a way that a good physical behavior 
for the potential is achieved. Examples for SU(3) gauge group can be found in reference
\cite{Deld}. The physical flux distribution leads to confinement 
of sources at intermediate distances and screening at large distances for zero 
n-laity representations. 

\section{Results and discussion}

\begin{figure*}
\center
\psfig{figure=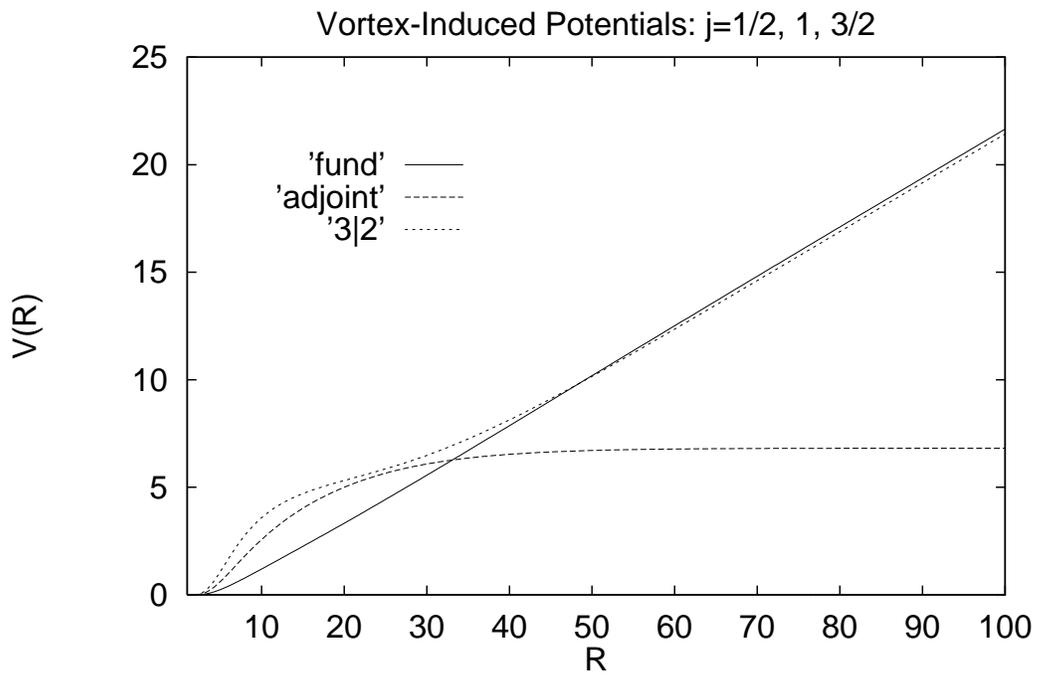,height=3.5in}
\caption{Interquark potential V(R) induced by center vortices for quarks
in the $j=\frac{1}{2}$, $1$, $\frac{3}{2}$ representations.}
\label{fig:su2}
\end{figure*}

\begin{figure*}
\center
\psfig{figure=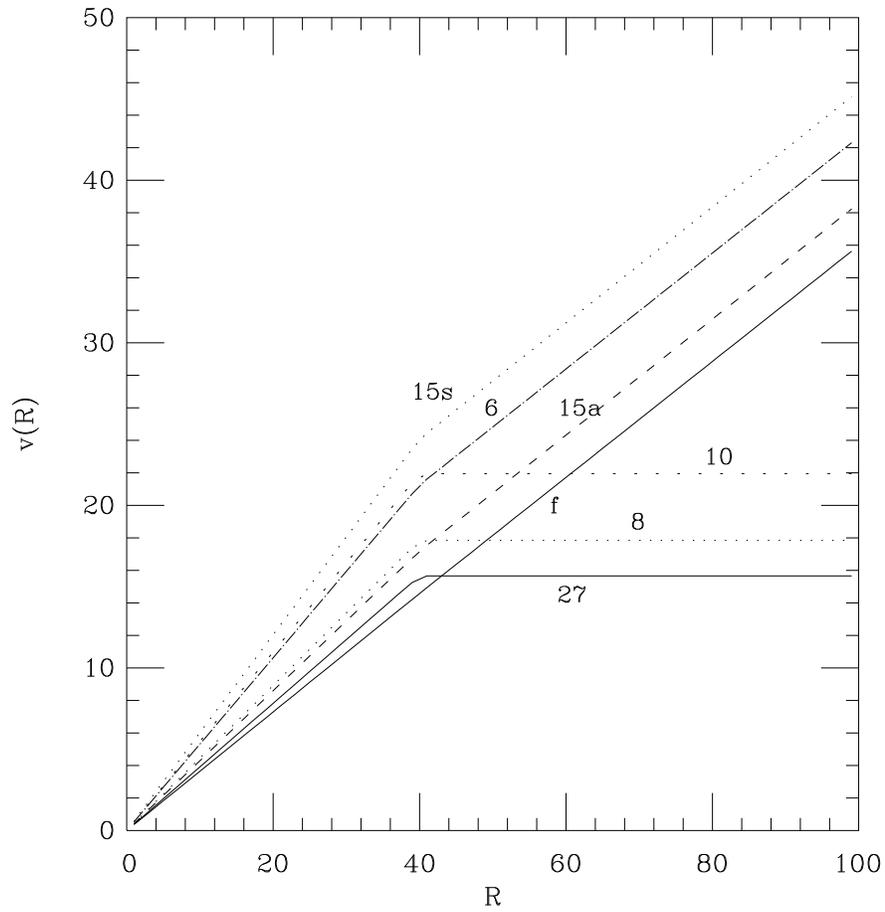,height=4.7in}
\caption{Potentials between static sources of SU(3). Zero triality representations
, 8, 27 and 10 are screened at large distances and others become parallel to
that of the fundamental representation. At intermediate distances $R \in [3,5]$
, the quark potentials are linear for all representations.}
\label{fig:su3}
\end{figure*}

\begin{figure*}
\center
\psfig{figure=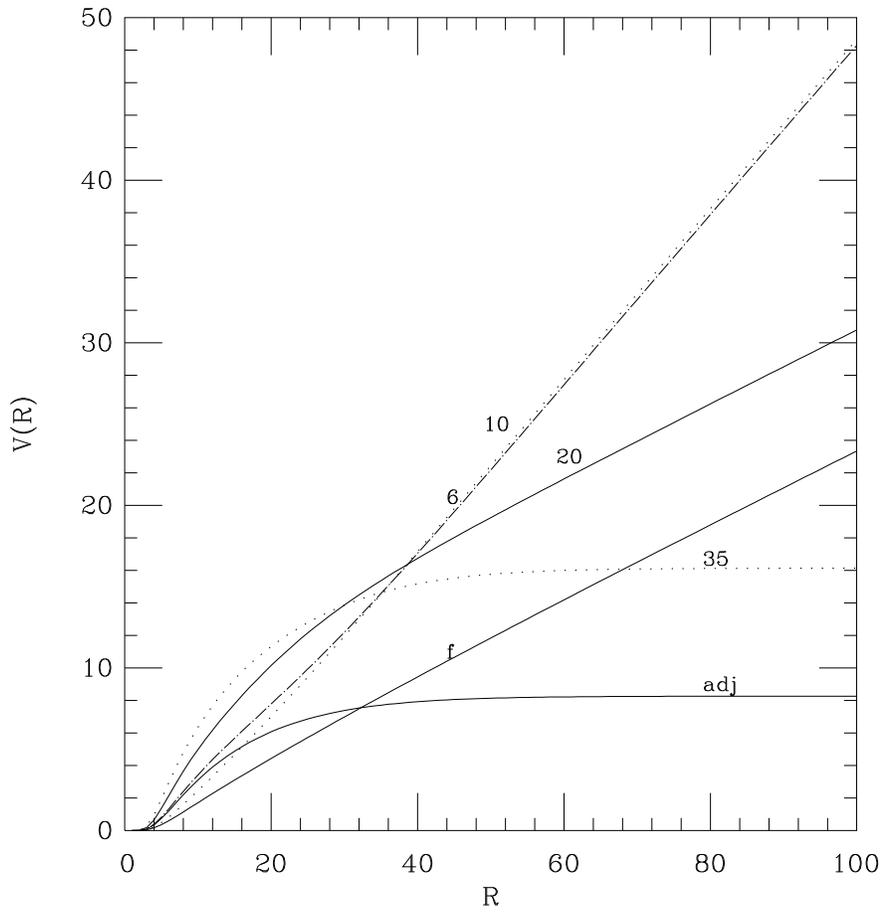,height=4.7in}
\caption{Potentials of static SU(4) sources. At large distances,
representations 15(adj) and 35 are screened; representation 20 gets the
same slope as fundamental representation and representation 10 is paralleled
to representation 6. String tensions at intermediate distances are 
qualitatively in agreement with the number of fundamental flux tubes.}
\label{fig:su4}
\end{figure*}

Figure \ref{fig:su2} shows the potential versus distance
for SU(2) quarks in the fundamental and representations with $j=1$
and $j=\frac{3}{2}$ \cite{Gren}. As seen from the plot, at large distances
adjoint quarks are screened and potential for quarks in the $j=\frac{3}{2}$
representation gets the same slop as that of the fundamental representation. 
This is because as the distance between quarks increases a pair of quarks in 
the adjoint representation (gluon-antigluon) pops out of the vacuum and makes
a pair of gluon-antigluon with initial adjoint quark-antiquark (gluon-antigluon)
such that the initial sources are not able to see each other (screened) and
the force between them is zero. For quarks in representation $\frac{3}{2}$
the gluon-antigluon pair which is created from the vacuum , interacts with the initial 
sources and the result is a pair of quark-antiquark in the fundamental
representation. That is why the slop of the potential changes to the slop
of that of the fundamental representation. This behavior is seen for SU(3)
\cite{Deld} and SU(4) \cite{Rafi} sources as well.
Figure \ref{fig:su3} shows the potential versus R for representations 6, 8, 10, 
15-symmetric, 15-antisymmetric and 27 in SU(3). At large distances, the force 
between quarks of zero-triality representations, 8, 10 and 27, is zero and for
non zero triality representations, 6, 15a and 15s, potentials get parallel to
that of the fundamental representation. Also for SU(4), figure \ref{fig:su4},
quarks in representations 15 and 35 are screened and non-zero 4-ality 
representations, 6, 10 and 20, become parallel to either representation
6 or 10. More details are available in reference \cite{Rafi}. On the other hand
at intermediate distances for all sources in all representations for all three 
gauge groups, there exists a linear potential. The string tension which is the 
coefficient of the linear term of the potential has an interesting property.   
Numerical calculations \cite{Bali} show that the string tension is proportional to the quadratic operator of the representation which is called Casimir scaling
\cite{Casm}. On the other hand it seems that 
the string tension is also proportion to the number of fundamental tubes 
embedded into the higher representation. 
This idea is called flux tube counting \cite{Shif}.
Our recent results especially for static sources in SU(4) agree better with flux tube counting 
than Casimir scaling. Tables 1 and 2 show the
results in SU(3) and SU(4), respectively. In each table, the first row 
indicates the representation. In the second row the number of original quarks
and anti-quarks, (n,m), anticipated in each representation is shown. The ratio
of Casimir scaling of each representation to the fundamental representation and
the ratio of string tensions are given in the third and fourth row, respectively.
The last row of each table indicates the number of fundamental fluxes exists
in each representation. The agreement between the ratio of string tensions
with both Casimir scaling and flux tube counting is qualitative but as seen
from  tables this agreement is better with flux tube counting than
Casimir scaling, especially for SU(4).

\begin{tablehere}
\setlength{\tabcolsep}{.1pc}
\caption{\label{tab1}This table shows Casimir numbers ratios, number of flux tubes and
string tensions ratios for SU(3) gauge group. Compare to Casimir scaling , a better agreement
between string tensions and flux tube counting is observed, especially for
higher representations.}
\begin{tabular}{|llcccccc|}
\hline
Repn.&  $3(fund.)$ & $8$ & $6$ & $15a$&  $10$ & $27$ & $15s$ \\\\
\hline
$(n,m)$    &  $(1,0)$ & $(1,1)$ & $(2,0)$ &  $(2,1)$ & $(3,0)$ & $(2,2)$ & $(4,0)$ \\ \\
$c_{r}/c_{f}$   &   $1$ &   $2.25$ &  $2.5$ &  $4$ &   $4.5$  &  $6.$ &  $7$ \\ \\
fund. fluxes &   $1$ &   $2$  &  $2$ &  $3$     &  $3$   &  $4$ & $4$   \\ \\
$k_{r}/k_{f}$   &    $1$ &  $2.02$ &    $2.21$ & $3.1$  &  $3.4$ &  $3.8$ &  $5.6$ \\ \\
\hline
\end{tabular}
\end{tablehere}

\begin{tablehere}
\setlength{\tabcolsep}{.12pc}
\caption{\label{tab2}The same as Table 1 but for quarks in SU(4).}
\begin{tabular}{|llccccc|}
\hline
Repn.&  $4(fund.)$ & $6$ & $15(adj.)$ & $10$&  $20$ & $35$ \\\\
\hline
$(n,m)$    &  $(1,0)$ & $(2,0)$ & $(1,1)$ &  $(2,0)$ & $(3,0)$ & $(4,0)$ \\ \\
$c_{r}/c_{f}$   &   $1$ &   $1.33$ &  $2.13$ &  $2.4$ &   $4.2$  &  $6.4$  \\ \\fund. fluxes &   $1$ &   $2$  &  $2$ &  $2$     &  $3$   &  $4$    \\ \\
$k_{r}/k_{f}$   &    $1$ &  $1.51$ &    $1.56$ & $1.76$  &  $2.31$ &  $2.66$ \\ \\
\hline
\end{tabular}
\end{tablehere}

\section*{Acknowledgments}
I would like to thank University of Tehran Research Council for support
of this work.

\end{multicols}
\end{document}